\documentclass[twocolumn,english,aps,superscriptaddress,prl,twocolumnm,nobalancelastpage]{revtex4-1}
\usepackage[T1]{fontenc}
\usepackage[latin9]{inputenc}
\usepackage{textcomp}
\usepackage{mathtools}
\usepackage{amsmath}
\usepackage{amssymb}
\usepackage{graphicx}
\usepackage{esint}

\makeatletter

\@ifundefined{textcolor}{}
{%
 \definecolor{BLACK}{gray}{0}
 \definecolor{WHITE}{gray}{1}
 \definecolor{RED}{rgb}{1,0,0}
 \definecolor{GREEN}{rgb}{0,1,0}
 \definecolor{BLUE}{rgb}{0,0,1}
 \definecolor{CYAN}{cmyk}{1,0,0,0}
 \definecolor{MAGENTA}{cmyk}{0,1,0,0}
 \definecolor{YELLOW}{cmyk}{0,0,1,0}
}

\makeatother

\usepackage{babel}
\begin{document}

\preprint{Draft \#1}

\title{Hydrodynamic and entropic effects on colloidal diffusion in corrugated channels}

\author{Xiang Yang}

\affiliation{Department of Physics and Astronomy and Institute of Natural Sciences,
Shanghai Jiao Tong University, Shanghai 200240, China}

\author{Chang Liu}

\affiliation{Department of Physics and Astronomy and Institute of Natural Sciences,
Shanghai Jiao Tong University, Shanghai 200240, China}

\author{Yunyun Li}

\affiliation{Center for Phononics and Thermal Energy Science, School of Physics
Science and Engineering, Tongji University, Shanghai 200092, China}

\author{Fabio Marchesoni }

\affiliation{Center for Phononics and Thermal Energy Science, School of Physics
Science and Engineering, Tongji University, Shanghai 200092, China}

\affiliation{Dipartimento di Fisica, Universit?i Camerino, I-62032 Camerino,
Italy}

\author{Peter Hänggi}

\affiliation{Institut für Physik, Universität Augsburg, D-86135 Augsburg, Germany}

\affiliation{Nanosystems Initiative Munich, Schellingstrasse 4, D-80799 München,
Germany}

\author{H. P. Zhang}

\email[]{ hepeng_zhang@sjtu.edu.cn}

\affiliation{Department of Physics and Astronomy and Institute of Natural Sciences,
Shanghai Jiao Tong University, Shanghai 200240, China}

\affiliation{Collaborative Innovation Center of Advanced Microstructures, Nanjing
210093, China}

\date{\today}
\begin{abstract}
In the absence of advection, confined diffusion characterizes transport in many natural and artificial
devices, such as ionic channels, zeolites, and nanopores. While extensive theoretical and 
numerical studies on this subject have produced many important predictions, experimental verifications of the predictions are rare. Here, we experimentally measure colloidal diffusion times in
microchannels with periodically varying width and contrast results
with predictions from the Fick-Jacobs theory and Brownian dynamics
simulation. While the theory and simulation correctly predict the
entropic effect of the varying channel width, they fail to account
for hydrodynamic effects, which include both an overall decrease and
a spatial variation of diffusivity in channels. Neglecting such hydrodynamic
effects, the theory and simulation underestimate the mean and standard
deviation of first passage times by 40\% in channels with a neck width
twice the particle diameter. We further show that the validity of
the Fick-Jakobs theory can be restored by reformulating it in terms
of the experimentally measured diffusivity. Our work thus demonstrates
that hydrodynamic effects play a key role in diffusive transport through
narrow channels and should be included in theoretical and numerical models.
\end{abstract}


\maketitle
Diffusive transport occurs prevalently in confined geometries \cite{Hanggi2009,Burada2009}.
Notable examples include the dispersion of tracers in permeable rocks
\cite{Berkowitz2006}, diffusion of chemicals in ramified matrices
\cite{Zeolites}, and the motion of submicron corpuscles in living
tissues \cite{Zhou2008,Bressloff2013}. Spatial confinement can fundamentally
change equilibrium and dynamical properties of a system by both limiting
the configuration space accessible to its diffusing components \cite{Hanggi2009}
and increasing the hydrodynamic drag \cite{Deen1987} on them. The
subject of confined diffusion is of paramount relevance to technological
applications and for this reason has been attracting growing interest
in the physics \cite{Hanggi2009,Burada2009}, mathematics \cite{Benichou2014},
engineering \cite{Berkowitz2006}, and biology communities \cite{Zhou2008,Hofling2013,Bressloff2013}.

From a theoretical point of view, particle diffusion in a confining
structure can be formulated in terms of a high dimensional Fokker-Planck
equation with appropriate boundary conditions reproducing the structure's
geometry. Such a boundary value problem is difficult to treat in general.
However, an approximate approach allows circumventing this difficulty
in the case of quasi-1D structures, such as ionic channels \cite{IonChannel},
zeolites \cite{Zeolites}, micro-fluidic channels \cite{Kettner2000,Matthias2003},
and nanopores \cite{Wanunu2010}. In these systems, transport takes
place along a preferred direction with the spatial constraints varying
along it. A typical example is represented by a corrugated narrow
channel. Focusing on the transport direction, Jacobs \cite{Jacobs1967}
and Zwanzig \cite{Zwanzig1992} assumed that the transverse degrees
of freedom equilibrate fast and proposed to eliminate them adiabatically
by means of an approximate perturbation scheme. In first order, they
derived a reduced diffusion equation, known as Fick-Jacobs (FJ) equation,
reminiscent of an ordinary 1D Fokker-Planck equation \textit{in vacuo},
except for two entropic terms, which locally modify the drift and
diffusion properties of the channeled particle \cite{Reguera2001,Kalinay2006,Reguera2006,Berezhkovskii2007,Burada2009}.

The FJ equation can be analytically solved to determine relevant transport
quantifiers, such as the effective mobility and diffusivity along
the channel. The theoretical predictions of the FJ reductionist approach
have been extensively checked against Brownian dynamics (BD) simulations,
which, on the contrary, propose to numerically integrate the full
multi-dimensional Langevin equation describing diffusion in arbitrary
geometries. Different types of channels have been investigated, including
2D \cite{Reguera2001,Burada2007,Berezhkovskii2015} and 3D channels
\cite{Ai2006,Berezhkovskii2007,Dagdug2011}, channels with abruptly
changing cross-sections \cite{Borromeo2010,Dagdug2011}, and curved
channels \cite{Bradley2009,Dagdug2012,Bauer2014}. On combining theoretical
and computational techniques, a variety of novel entropy-driven transport
mechanisms have been predicted, such as drive-dependent mobilities
\cite{Burada2009,Reguera2006,Burada2007}, stochastic resonance \cite{Burada2008,Ding2015},
absolute negative mobilities \cite{Haenggi2010}, entropic rectification
\cite{Marchesoni2009,Malgaretti2013}, and particle separation \cite{Reguera2012}.
Several of these results are presently recognized as being of both
fundamental and technological importance.

Surprisingly, experimental studies of entropic effects on confined
diffusion are still scarce \cite{Marquet2002,Matthias2003,Huang2004,Mathwig2011},
mainly due to technical difficulties encountered in fabricating micron-sized
corrugated channels of controlled width \cite{Pagliara2014}. Here,
we implemented a two-photon direct laser writing technique to overcome
this experimentally difficult problem and fabricated channels with
systematically modulated cross-sections. We then measured the diffusive
dynamics of micrometric colloidal particles through such channels
by standard video microscopy and compared the outcome with predictions
obtained by FJ approximation and from BD simulation. We discover that,
as the channel's width shrinks towards the particle's diameter, hydrodynamic
effects \cite{Deen1987,HappelBook,Volpe2010,Chen2011a,Dettmer2014,Skaug2015},
largely ignored in previous studies, grow in strength and become comparable
to the predicted entropic effects, thus indicating an unexpected breakdown
of the standard FJ theory and BD simulation in narrow channels. We
further show that hydrodynamic effects can be incorporated by using
an experimentally measured local diffusivity. With such a phenomenological
modification, the FJ theory and BD simulation accurately predict the
experimental data.

\section*{Results}

\subsection*{Experimental realization}

\begin{figure}
\centering{}\includegraphics[width=8cm]{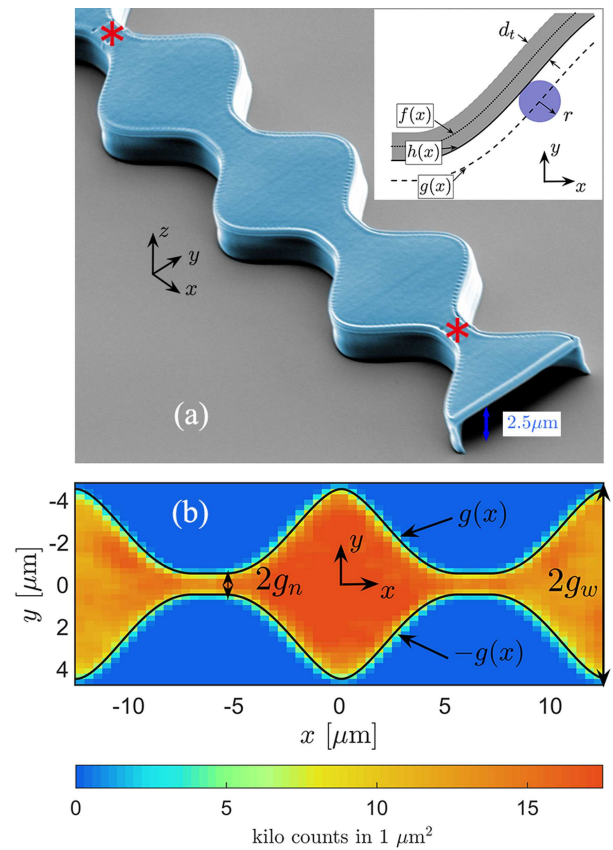}\protect\protect\caption{\textbf{| Channel geometry.} (\textbf{a}) Scanning electron image
of a channel of inner height $2.5\mu$m. Narrow openings at the two
ends are marked by red symbols. The inset illustrates the channel's
geometry: the laser scanning contour, $f(x)$, the wall inner boundary,
$h(x)$, and the effective boundary of the space accessible to the
particle center, $g(x)$; $d_{t}\simeq0.7\mu$m and $r=0.5\mu$m are
the wall thickness and the particle radius, respectively. (\textbf{b})
Spatial distribution of particle counts in a typical 20-hour-long
experiment. The effective channel boundary is marked by black lines
and is denoted by $\pm g\left(x\right)$, see Eq. (\ref{E1}); here,
$g_{n}=0.5\mu$m, and $g_{w}=4.5\mu$m. }

\label{F1} 
\end{figure}

Our channels were fabricated on a cover slip by means of a two-photon
direct laser writing system, which solidifies polymers according to
the preassigned channel profile, $f(x)$, with a sub-micron resolution.
As shown in Fig. \ref{F1}, the quasi-two-dimensional channel has
a uniform height of $2.5\mbox{}\mu$m ($z$ direction). The curved
side walls are $0.7\mu$m thick and their inner side walls a distance
$\pm h(x)$ away from the channel's axis ($x$ direction). 

After fabrication, channels were immersed in water with fluorescently-labeled
Polystyrene spheres of radius $r=0.5\mu$m. A holographic optical
tweezer was used to drag a particle into the channel through a narrow
entrance {[}red symbols in Fig. \ref{F1}(a){]}. The entrances are
barely wider than the particle diameter so as to create insurmountable
entropic barriers \cite{Burada2009}, which prevent the particle inside
the channel from escaping. Particle motion in the quasi-two-dimensional
channel was recorded through a microscope at rate of 30 frame/sec
for up to 20 hours. The projected particle trajectory in $xy$ plane
was extracted from the recorded videos by standard particle tracking
algorithms; particle diffusion perpendicular to the imaging $xy$
plane was not resolved in our measurements. 

\begin{figure*}
\centering{}\includegraphics[width=1\textwidth]{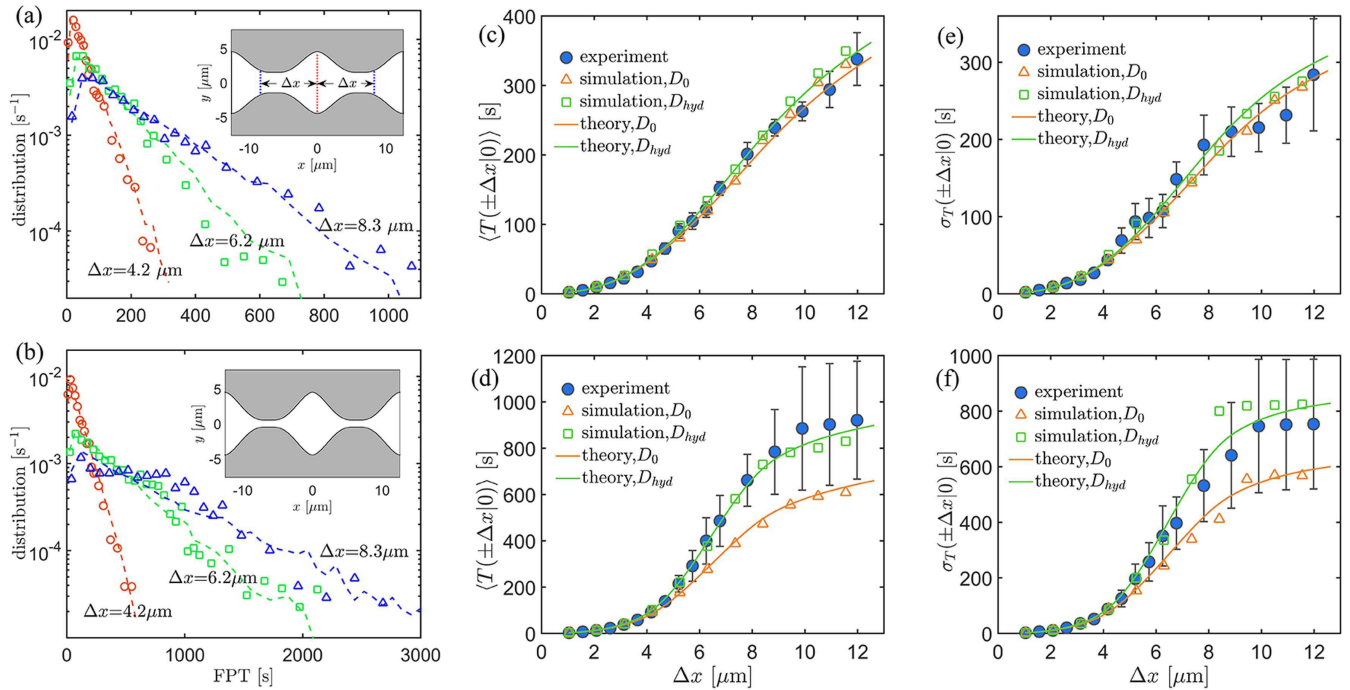}\protect\protect\caption{\textbf{| Statistics of the first passage times.} Probability distributions,
(\textbf{a}, \textbf{b}), averages, (\textbf{c}, \textbf{d}), and
standard deviations, (\textbf{e}, \textbf{f}) of the FPT's in a wide
(top row, $g_{n}=1.5\mu$m) and narrow (bottom row, $g_{n}=0.5\mu$m)
channel with same maximum half-width, $g_{w}=4.5\mu$m. In (\textbf{a},
\textbf{b}), experimental data (symbols) are compared with the outcome
of Brownian dynamics simulations (curves) for the spatially modulated
diffusivity, $D_{hyd}(x)$. The relevant channel profiles, $\pm g(x)$,
are shown in the insets of (\textbf{a}, \textbf{b}); vertical dashed
segments mark the starting (red, $x=0$) and ending (blue, $x=\pm\Delta x$)
positions of the first passage events. In (\textbf{c}-\textbf{f}),
experimental, numerical, and theoretical results are shown as solid
symbols, empty symbols, and solid curves, respectively. Numerical
and theoretical results with constant, $D_{0}$, and varying, $D_{hyd}$,
diffusivity are color coded, respectively, in orange and green. \label{F2}}
\end{figure*}

Inside the channels, the particle diffuses in a flat energy landscape.
To show that, we quantized the measured particle coordinates $(x,y)$
into small bins (0.4$\times$0.25 $\mu\mbox{m}^{2}$) and counted
the number of times the particle enters each bin. As shown in Fig.
\ref{F1}(b), particle counts are uniformly distributed with a standard
variation about 12\% of the mean. Regions where the particle counts
drop sharply to zero are inaccessible to the particle's center and,
in Fig. \ref{F1}(b), are delimited by the black curves {[}see also
the inset of Fig. \ref{F2}(a){]}. The effective channel's boundary
{[}denoted by $g(x)${]} is a periodic function; in the central region,
the boundary was given the form of a cosine, which then tapers off
to a constant in correspondence with the bottlenecks, that is

\begin{equation}
g(x)=\begin{cases}
\frac{1}{2}\left(g_{w}+g{}_{n}\right)+\frac{1}{2}\left(g_{w}-g{}_{n}\right)\cos(\frac{16\pi x}{7L}),\mbox{ }|x|<\frac{7}{16}L\\
g_{n},\mbox{ \hspace{1cm}}\frac{7}{16}L\leq|x|\leq\frac{L}{2} & \negmedspace
\end{cases}\label{E1}
\end{equation}
The length of the channel unit cell was kept fixed in all experiments,
$L=12.5\mu$m, while the parameters $g_{n}$ and $g_{w}$, representing
respectively, its minimum and maximum half-width, were varied. For
the channel shown in Fig. \ref{F1}(b), $g_{n}=0.5\mu$m, and $g_{w}=4.5\mu$m.

\subsection*{First passage time statistics}

A direct measurement of the diffusion constant in channels \cite{Hanggi2009},
$D_{c}=\lim_{t\to\infty}\langle[x(t)-x(0)]^{2}\rangle/\left(2t\right)$,
would require fabricating a much longer (linear or circular) channel
structure. As a more viable alternative we measured the First Passage
Times (FPT) \cite{Goel1974,RevModPhys.62.251,Benichou2014}. As in
the FJ theory, we focus on the particle motion along the channel direction
and measure the duration of the unconditional first passage events
that start at $x=0$ {[}red segment in the inset of Fig. \ref{F2}(a){]}
and end at $x=\pm\Delta x$ (blue segments), with no restriction on
the transverse coordinate $y$. Distributions of experimentally measured
unconditional FPTs, also denoted by $T(\pm\Delta x|0)$, are plotted
in Figs. \ref{F2}(a,b); all distributions (for three $\Delta x$
values in two channels of different bottleneck half-width, $g_{n}$)
exhibit an exponential tail, similar in spirit with the narrow-escape
problem \cite{Benichou2014}. From these measured FPT distributions,
we extract the means, $\left\langle T(\pm\Delta x|0)\right\rangle $,
and the standard deviations, $\sigma_{T}(\pm\Delta x|0)\coloneqq\sqrt{\langle T^{2}(\pm\Delta x|0)\rangle-\langle T(\pm\Delta x|0)\rangle^{2}}$;
our results are plotted in Figs. \ref{F2}(c-f) against the diffusing
distance, $\Delta x$. A decrease of the bottleneck width, $g_{n}$,
from $1.5\mu$m in (c) to $0.5\mu$m in (d), sharply increases the
diffusion time. For instance, the mean FPT to the center of the adjacent
cells, $\left\langle T(\pm L|0)\right\rangle $, nearly triples from
300s in (c) to 900s in (d). A similar increase can be observed in
the standard deviations, $\sigma_{T}(\pm\Delta x|0)$, depicted in
Figs. \ref{F2}(e,f). To this regard, we notice that, for both channels,
the experimental curves $\langle T(\pm\Delta x|0)\rangle$ and $\sigma_{T}(\pm\Delta x|0)$,
almost overlap, as to be expected in view of the exponential decay
of the relevant FPT distributions \cite{RevModPhys.62.251}. Accordingly,
the corresponding channel diffusion constant, $D_{c}$, can be estimated
in terms of an appropriate mean FPT; that is \cite{RevModPhys.62.251},
$D_{c}=L^{2}/\left(2\langle T(\pm L|0)\rangle\right)$.

We next compare our experimental data with the predictions of the
standard FJ theory and BD simulations. The channel geometry renders
our experimental system effectively two-dimensional; analytical and
numerical studies were carried out in the same dimension. Following
the FJ scheme and taking advantage of symmetry properties of our experiments,
we calculate the analytical expression, 
\begin{equation}
\langle T_{FJ}(\pm\Delta x|0)\rangle=\int_{0}^{\Delta x}\frac{d\eta}{g(\eta)\mathbb{D}(\eta)}\int_{0}^{\eta}g(\xi)d\xi,\label{E2}
\end{equation}
for the mean FPT. Here, $\mathbb{D}\left(x\right)$ is the effective
local diffusivity containing the entropic corrections that result
from the adiabatic elimination of the transverse coordinate, $y$.
Among the (slightly) different functions $\mathbb{D}\left(x\right)$
proposed in the recent literature \cite{Berezhkovskii2015}, we adopted
the Reguera-Rub\`{i} heuristic expression \cite{Reguera2001}, i.e.,
\begin{equation}
\mathbb{D}(x)=\frac{D_{0}}{[1+g'(x)^{2}]^{1/3}},\label{E3}
\end{equation}
where $g'(x)$ is the slope of the channel's profile $g(x)$, and
$D_{0}$ is the particle's diffusivity away from side walls. We also
calculated the second FPT moment, 
\begin{equation}
\langle T_{FJ}^{2}(\pm\Delta x|0)\rangle=\int_{0}^{\Delta x}\frac{2d\eta}{g(\eta)\mathbb{D}(\eta)}\int_{0}^{\eta}g(\xi)\langle T_{FJ}(\pm\Delta x|\xi)\rangle d\xi,\label{E4}
\end{equation}
where $\langle T_{FJ}(\pm\Delta x|\xi)\rangle$ reads like in Eq.
(\ref{E2}), except that the outer integral runs here from $\xi$
to $\Delta x$. The derivation of Eqs. (\ref{E2}) and (\ref{E4})
can be found in the Supplementary Information (SI).

\begin{figure}
\centering{}\includegraphics[width=8cm]{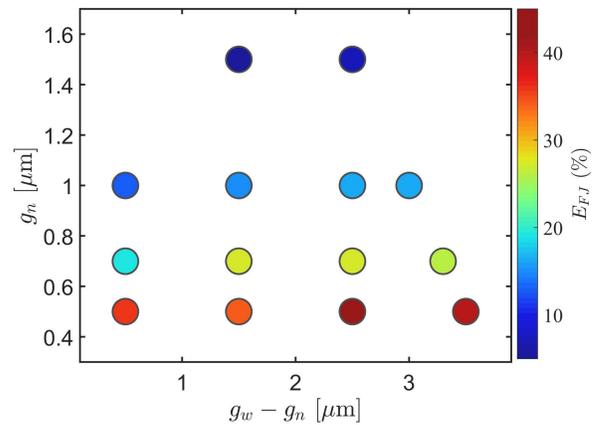}\protect\protect\caption{\textbf{| Deviation of theoretical predictions from experimental results.
}Relative deviation, $E_{FJ}$, increases with decreasing the bottleneck
half-width, $g_{n}$, and weakly depends on the modulation amplitude,
$g_{w}-g_{n}$, of the channel. \label{F3}}
\end{figure}

To use Eqs. (\ref{E2}) to (\ref{E4}), we need to know the diffusivity,
$D_{0}$. Unlike in an unbounded space, where the diffusivity of a
sphere is determined by the Stokes-Einstein equation, there is no
general expression for the diffusivity of a colloidal particle in
a confined geometry. Hence, we experimentally measured $D_{0}$ by
monitoring the diffusion of the particle about the center of a channel's
cell, where the entropic effects are minimal, and for displacements
smaller than one particle radius. The FJ expressions (\ref{E2}) and
(\ref{E4}) were then computed explicitly for the measured value of
$D_{0}$ and the actual channel geometry (namely, the parameters $g_{n}$,
$g_{w}$, and $L$). For the sake of a comparison, 2D BD simulations
were also performed for the same model parameters. Theoretical and
numerical results, orange symbols and curves in Fig. \ref{F2}(c-f),
agree closely with each other for both the wide and narrow channel.
The comparison with the experimental data, instead, is satisfactory
only in the case of the wide channel, Figs. \ref{F2}(c, e). For the
narrow channel of Figs. \ref{F2} (d, f), the experimental data with
$\Delta x>6\mu\mbox{m}$ are as much as 40\% larger than the predicted
values. To further investigate this discrepancy, we carried out experiments
in channels with different width parameters, $g_{w}$ and $g_{n}$;
the discrepancy is quantified in Fig. \ref{F3} by the relative mean-FPT
difference at $\Delta x=L/2$ (bottleneck midpoints), $E_{FJ}={[\left\langle T(\pm L/2|0)\right\rangle -\left\langle T_{FJ}(\pm L/2|0)\right\rangle ]}/{\left\langle T_{FJ}(\pm L/2|0)\right\rangle }$.
For narrow channels, the experimental data are consistently larger
than the corresponding theoretical and numerical predictions. The
discrepancy depends weakly on the amplitude of the channel modulation,
$g_{w}-g_{n}$, but increases significantly with decreasing bottleneck
half-width, $g_{n}$.

\subsection*{Diffusivity measurements}

The theoretical and numerical predictions discussed so far assume
a constant particle diffusivity, $D_{0}$, throughout the channel,
which is a reasonable approximation for particle diameters much smaller
than the channel width. However, this assumption is doomed to fail
for small bottleneck widths (when the FJ approach is supposed to work
best), because the proximity of no-slip side walls in the neck regions
is known to increase the hydrodynamic drag on a finite-size particle
and, therefore, suppress its local diffusivity \cite{Eral2010,Volpe2010,Cervantes-Martinez2011,Chen2011a,Dettmer2014,Skaug2015}.

\begin{figure}
\centering{}\includegraphics[width=8cm]{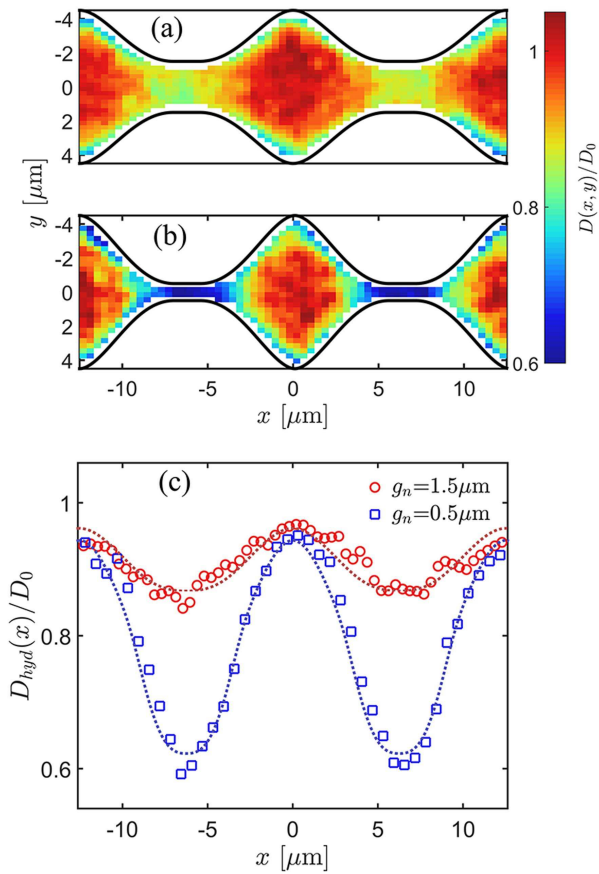}\protect\protect\caption{\textbf{| Spatial variability of the local diffusivity.} (\textbf{a},
\textbf{b}) Normalized diffusivity from experiments, $D\left(x,y\right)/D_{0}$,
measured in the wide and narrow channels of Fig. \ref{F2}. (\textbf{c})
Normalized diffusivity, $D_{hyd}\left(x\right)/D_{0}$, versus $x$.
Experimental data and results from finite-element calculations are
represented respectively by symbols and curves. Experimentally measured
$D_{0}$ is $0.29$ $\mu$m$^{2}$/s and $0.27$ $\mu$m$^{2}$/s
for wide and narrow channels, respectively; the Stokes-Einstein equation
predicts a sphere diffusivity of $0.5$ $\mu$m$^{2}$/s in an unbounded
space. \label{F4}}
\end{figure}

To demonstrate such a hydrodynamic effect in our device, we measured
the particle diffusivity inside the channel. At any given location,
$(x,y)$, we recorded the particle mean-squared displacement in the
$x$ direction, $\left\langle \delta x^{2}\left(x,y\right)\right\rangle $,
for a time interval $\delta t=0.2$s and estimated the local diffusivity
through Einstein's law, $D(x,y)=\left\langle \delta x^{2}\left(x,y\right)\right\rangle /\left(2\delta t\right)$.
As shown in SI, the value chosen for $\delta t$ is long enough to
ensure a good statistics for our measurements of $D(x,y)$, but not
enough for the entropic effects and spatial variability of $D(x,y)$
due to the channel modulation to become detectable. Measurements of
$D(x,y)$ in the wide and narrow channels are shown in Fig. \ref{F4}(a,
b). In both, $D(x,y)$ is largest in the open regions at the center
of the unit cells, and strongly suppressed in the bottlenecks. In
the spirit of the FJ theory, we average $D(x,y)$ along the transverse
direction 
\begin{equation}
D_{hyd}\left(x\right)\coloneqq\frac{1}{2g(x)}\int_{-g(x)}^{g(x)}D(x,y)dy
\end{equation}
 and plot $D_{hyd}\left(x\right)/D_{0}$ as a function of $x$ in
Fig. \ref{F4}(c). The spatial variability of $D_{hyd}\left(x\right)$
is about 10\% and 40\% for the wide and narrow channels, respectively.

We corroborate the local diffusivity measurements with full hydrodynamic
computations. The hydrodynamic friction coefficient in the $x$ direction
$\gamma(x,y)$ was computed by means of a finite-element package (COMSOL);
the local diffusivity was calculated via the fluctuation-dissipation
theorem, $D(x,y)={k_{B}T}/{\gamma(x,y)}$, and the result is averaged
over the $y$ coordinate to obtain $D_{hyd}\left(x\right)$. Results
from finite-element calculations are shown in Fig. \ref{F4}(c) as
curves and are in excellent agreement with the experimental data.

\subsection*{Hydrodynamic correction}

Figure \ref{F4} depicts that particle diffusion through narrow bottlenecks
can be significantly slower than in the wide region; moreover, the
spatial modulation of the local particle diffusivity increases with
decreasing the bottleneck width. This is in clear contrast with the
assumption of constant diffusivity we adopted above, when implementing
the FJ-formalism and the BD simulation code. To appreciate the effect
of the spatial dependence of the local diffusivity, we replace the
constant diffusivity, $D_{0}$, with the experimental measurement,
$D_{hyd}\left(x\right)$, reported in Fig. \ref{F4}(c), both in the
theoretical treatment and in the numerical code. The new analytical
and numerical predictions are in plotted Fig. \ref{F2}(c-f), respectively,
as green curves and symbols. Their agreement with the experimental
data is excellent. Furthermore, we used the improved BD code to compute,
besides the first two FPT moments, also the FPT distributions displayed
in Figs. \ref{F2}(a, b). Again, the close comparison obtained with
the experimental data confirms the validity of our phenomenological
approach.

\section*{discussions}

The coincidence of approximate analytical predictions and simulation
results occurs for any choice of the local diffusivity, i.e., $D_{0}$
or $D_{hyd}\left(x\right)$, as illustrated in Fig. \ref{F2}. This
means that the FJ theory well describes the entropic effects of particle
transport in weakly corrugated channels with $\left|g'(x)\right|<1$
\cite{Berezhkovskii2015}. However, assuming constant particle diffusivity,
as common practice in the current literature, can lead to large discrepancies
between theoretical predictions and experimental observations. Indeed,
to correctly analyze the diffusion of finite-size particles in narrow
channels one needs to account for the hydrodynamic effects, as well.
Because there is no general analytical solution for particle diffusivity
in a corrugated confinement, we substituted the constant diffusivity,
$D_{0}$, with an empirical function from experimental measurements,
$D_{hyd}(x)$. The substitution $D_{0}\to D_{hyd}(x)$ in Eq. (\ref{E3}),
suggests a phenomenological factorization of entropic and hydrodynamic
effects, whose validity is justified \textit{a posteriori} by the
reported close comparison with the experimental data. The FJ theory
with hydrodynamic corrections thus remains a powerful analytical tool
to investigate diffusion in complex channels.

The local diffusivity in corrugated channels displays a rich 2D structure,
see Fig. \ref{F4}. The comparison with a more tractable geometry
helps illustrating the phenomenon of the hydrodynamic diffusivity
suppression advocated above. For a spherical particle of a radius
$r$ diffusing along the axis of a relatively long cylinder \cite{HappelBook,Misiunas2015},
the particle diffusivity is approximated by, 
\begin{equation}
D\approx D_{u}\left[1-2.104\left(\frac{r}{R}\right)+2.089\left(\frac{r}{R}\right)^{3}\right],\label{E5}
\end{equation}
where $D_{u}$ is the particle diffusivity in an unbounded space and
$R$ denotes the cylinder radius. According to Eq. (\ref{E5}), particle
diffusivity in confined geometries is generally smaller than in an
unbounded space. In our channels, the maximum diffusivity $D_{0}$
is about 60\% of the Stokes-Einstein predicted value. Diffusivity
also tends to decrease as the confinement grows tighter (i.e., for
larger $r/R$); this explains why diffusivity is smaller in the necking
regions of our channel. In certain applications, such as the entropic
splitters \cite{Reguera2012}, one has recourse to tight confinement
to generate high entropic barriers; we expect hydrodynamically suppressed
diffusion to play an important role in these situations and possibly
boost the separation efficiency.

In most technological applications, particles are driven by external
fields. This may prevent the system from equilibrating in the transverse
directions and produce new, more complex transport mechanisms \cite{Burada2007}.
Hydrodynamics also plays a more active role in the presence of external
driving \cite{Martens2013a}. Our experimental system can serve as
an excellent platform to investigate these important and challenging
problems.

\subsection*{Methods }

\subsection*{{\small{}{}Channel fabrication and Imaging procedure}}

{\small{}{}Microchannels were fabricated with a two-photon direct
laser writing system ($\mu$FAB3D from Teem Photonics). This system
uses a microscope objective lens (Zeiss Fluar 100$\times$, numerical
aperture 1.3) to focus pulsed laser (Nd:YAG microchip laser with 532nm
wavelength, 750ps pulse width, and 40kHz repetition rate) into a droplet
of photoresist resin that is mounted on a piezo-nanopositioning stage
(PI P-563.3CL). We used a polymer resin ORMOCOMP (Micro resist technology,
GmbH) with a photo-initiator (1,3,5-Tris(2-(9-ethylcabazyl-3)ethylene)benzene).
Photopolymerization occurs and solidifies the resin at the focal point;
the piezo-stage scans the resin relative to the focal point along
a preassigned trajectory {[}$f\left(x\right)$ in the inset of Fig.
1(a){]} to fabricate the desired structure. After the scanning is
finished, the remaining liquid resin was removed by washing the structure
with 4-Methyl-2-pentanone and then acetone for 5 minutes. Then channels
were thoroughly cleaned with distilled water to prevent particles
from sticking to the channel boundaries.}{\small \par}

{\small{}Fluorescently-labeled Polystyrene particles were purchased
from Invitrogen (Catalog number: F13080). Particle motion was recorded
through a 60\texttimes{} oil objective (numerical aperture 1.3) in
an inverted fluorescent microscope (Nikon Ti-E). With the help of
an autofocus function (Nikon perfect focus), we imaged the diffusion
of a colloidal particle in the channel for up to 20 hours at room
temperature ( 27 $^{\circ}\mbox{C}$).}{\small \par}

\subsection*{{\small{}{}Brownian dynamics simulation}}

{\small{}{}The motion of a colloid particle is governed by a 2D overdamped
Langevin equation in simulations. The particle diffusivity varies
spatially when the diffusivity function $D_{hyd}(x)$ is used; for
thermodynamic consistency, we adopted the transport (also known as
kinetic or isothermal) convention \cite{Hanggi1978,Sokolov2010a,Farago2014a,Bruti-Liberati2008a}
to compute the stochastic integral \cite{Hanggi1978,vanKampen1981}.
{}The channel boundary was represented by a string of fixed particles,
which interact with the colloidal particle via a short-range repulsive
force. Particle trajectories from simulation were analyzed in the
same way as their experimental counterpart to extract the effective
volume of the channel's unit cell and the FPT's. See SI for more details.}{\small \par}

\subsection*{{\small{}{}Finite-element calculation }}

{\small{}{}We solved the Stokes equations in a typical setup shown
in Fig. S2(a). No-slip boundary conditions were imposed on the side
walls, floor and ceiling, and open boundary conditions at the channel
openings. The geometry of the side wall was set to reproduce the inner
channel boundary measured in the experiments {[}see insert of Fig.
1(a){]}. A sphere was driven with a constant speed, $v_{x}$, in the
$x$ direction; at different points, $(x,y)$, on a horizontal plane.
We measured the drag force, $f_{x}$, and computed the hydrodynamic
drag coefficient, $\gamma(x,y)=f_{x}/v_{x}$. See SI for more details.}{\small \par}

\end{document}